\documentclass[fleqn,10pt]{wlscirep}
\usepackage[utf8]{inputenc}
\usepackage[T1]{fontenc}
\usepackage{bm}
\usepackage{graphicx}
\usepackage{xcolor}
\usepackage{hyperref}
\usepackage{caption}
\usepackage{subfig}
\usepackage{booktabs}
\usepackage{amsmath,amsfonts}
\usepackage{multirow}
\usepackage{makecell}
\usepackage{pifont}
\usepackage[table]{xcolor}
\usepackage{upgreek}
\graphicspath{{figs/}}

\usepackage{soul} 
\usepackage{xcolor} 
\usepackage{xcolor}
\definecolor{wordyellow}{RGB}{255, 255, 0}  
\definecolor{lightgray}{gray}{0.9}
\sethlcolor{wordyellow}
\soulregister{\cite}{1}

\usepackage[justification=justified,singlelinecheck=false]{caption}

\usepackage{float}

\title{Passive All-Optical Nonlinear Neuron Activation via PPLN Nanophotonic Waveguides}

\author[1,2,$^{\dagger}$]{Wujie Fu}
\author[3,4,$^{\dagger}$]{Xiaodong Shi}
\author[5]{Sakthi Sanjeev Mohanraj}
\author[1]{Lei Shi}
\author[1]{Yuan Gao}
\author[1]{Zexian Wang}
\author[1]{Jianing Wang}
\author[5]{Xu Chen}
\author[1]{Luo Qi}
\author[1]{Pragati Aashna}
\author[6,7]{Guanyu Chen}
\author[1,3,4,5,*]{Di Zhu}
\author[1,2,*]{Aaron Danner}
\affil[1]{Department of Electrical and Computer Engineering, National University of Singapore, 117583, Singapore}
\affil[2]{Graduate School for Integrative Sciences and Engineering, National University of Singapore, 119077, Singapore}
\affil[3]{A$^\ast$STAR Quantum Innovation Centre (Q.InC), Agency for Science, Technology and Research (A$^\ast$STAR), 138634, Singapore}
\affil[4]{Institute of Materials Research and Engineering (IMRE), Agency for Science, Technology and Research (A$^\ast$STAR), 138634, Singapore}
\affil[5]{Department of Materials Science and Engineering, National University of Singapore, 117575, Singapore}
\affil[6]{Key Laboratory of Optoelectronic Technology \& Systems Ministry of Education, Chongqing University, Chongqing 400044, China}
\affil[7]{College of Optoelectronic Engineering, Chongqing University, Chongqing 400044, China}
\affil[$^{\dagger}$]{These authors contributed equally.}
\affil[*]{Corresponding authors: adanner@nus.edu.sg; dizhu@nus.edu.sg}

\begin{abstract}
Artificial intelligence (AI) is transforming modern life, yet the growing scale of AI applications places mounting demands on computational resources, raising sustainability concerns. Photonic integrated circuits (PICs) offer a promising  alternative, enabling massive parallelism, low latency, and reduced electrical overhead, particularly excelling in high-throughput linear operations. However, passive and fully optical nonlinear activation functions with equally superb performance remain rare, posing a critical bottleneck in realizing all-optical neural networks in PICs. 
Here, we demonstrate a compact and integrated all-optical nonlinear activation method, experimentally realized through strong second-order optical nonlinearities in periodically poled lithium niobate (PPLN) nanophotonic waveguides, achieving $\sim$80\% absolute conversion efficiency. This activation exhibits a sigmoid-like, wavelength-selective response with femtosecond-scale dynamics and light-speed processing, operating passively without external control and auxiliary signals.
We validate its feasibility for neural inference by cascading the PPLN-driven activations with a linear silicon PIC, demonstrating all-optical nonlinear neuron expressivity. Moreover, combining the measured nonlinearity with linear operations calculated by the PIC, we show that PPLN-activated multi-layer optical neural networks can achieve performance on par with digital implementations in real-world tasks, including airfoil regression and medical image classification.
These results pave the way toward scalable, high-speed, and fully integrated all-optical neural networks for next-generation photonic AI hardware.

\end{abstract}

\begin{document}

\flushbottom
\maketitle

\thispagestyle{empty}

\section*{Introduction}
Artificial general intelligence (AGI) has emerged as a widely pursued goal, showing the potential to revolutionize diverse fields such as climate prediction \cite{1}, autonomous driving \cite{2}, drug discovery \cite{3}, controlled nuclear fusion \cite{101}, and other areas of scientific research and engineering \cite{102}.
The accelerating progress toward AGI is driven by two key enablers: the continued advancement of high-performance computing hardware, which offer high-throughput computational power \cite{6}; and the rapid evolution of frontier AI models like large language models, which leverage nonlinear dynamics to learn complex, high-dimensional representations from vast datasets \cite{4}.
However, the unprecedented scale and complexity of modern AI models are pushing the limits of conventional electronic computing architectures imposed by Moore's law \cite{8}. As model sizes reach hundreds of billions of parameters, the resulting computational demands present significant challenges in terms of bandwidth \cite{109}, latency \cite{103}, and energy efficiency \cite{104}.
These limitations have sparked growing interest in alternative computing paradigms that can natively support the massive parallelism and data-intensive demands of increasing AI workloads. Optical computing, particularly in the form of optical neural networks (ONNs), has emerged as a promising candidate, offering intrinsic advantages such as ultrafast signal transmission \cite{9}, passive data transport without resistive losses \cite{105}, and capability for low-power, high-speed computation \cite{5}.

Within the landscape of optical computing, photonic integrated circuits (PICs) stand out for their scalability and manufacturability \cite{hua2025pace}. By enabling dense integration of photonic components within low-loss waveguide structures, PICs support compact, foundry-ready manufacturable architectures and provide a practical pathway toward large-scale, chip-level deployment of ONNs \cite{67}.
Mathematically, ONNs are built upon two core computational primitives: linear transformations, commonly realized as optical matrix multiplications, and nonlinear activations, which are essential for expressing complex decision boundaries and latent feature representations.
Linear operations are fundamental to feature embedding and signal transformation, and have been successfully demonstrated on PIC platforms using various photonic architectures.
Examples include Mach–Zehnder interferometers (MZIs) \cite{11,28,yan2025deep}, wavelength-division multiplexing (WDM) \cite{30,he2025LY,yu2025wdm}, optical metasurfaces \cite{32}, and phase-change materials \cite{22,108}. These advances highlight the growing maturity and strong potential of PIC-based systems for large-scale optical computing, approaching and in some regimes exceeding the capabilities of electronic processors \cite{10}.

In contrast, realising optical nonlinear neuron activations with computational performance rivaling their linear counterparts remains a critical challenge for ONNs on PIC platforms \cite{105}.
Currently, most ONNs rely on incoherent optoelectronic approaches for nonlinear activation, which involve optical-to-electrical conversion to detect optical power \cite{28,zheng2024MZI_circle}. This process inevitably introduces latency and power inefficiencies and typically restricts activations to positive-only nonlinearities \cite{23}.
To fully exploit the computational superiority of photonics, various optical-to-optical activation methods have been explored, including material absorption \cite{44,39,70,107}, laser bistability \cite{36}, semiconductor optical amplifiers (SOAs) \cite{41}, stimulated Brillouin scattering (SBS) \cite{43}, microring resonators (MRRs) \cite{42,52,gongqh2025to}, parametric processes \cite{51,73,li2024ppln}, and exciton polariton \cite{gan2025ep}. 
Among these, $\chi^{(2)}/\chi^{(3)}$ nonlinearities are fundamentally attractive owing to their coherent and near-instantaneous response, opening a window for ultrafast operation. 
Bulk-optics demonstrations have validated parametric processes for nonlinear neurons \cite{68,li2024ppln}, yet remain hampered by modest efficiency, losses, and thermal–mechanical sensitivities.
Meanwhile, most existing approaches rely on external active control—such as electrical driving, thermal tuning, or auxiliary optical inputs—to initiate or stabilize their nonlinear response \cite{36,51,52,gongqh2025to,73,43,41,70,107,li2024ppln}. These additional requirements introduce extra system overhead and routing complexity, which can constrain the achievable speed and overall throughput of photonic computing systems.
Therefore, a fully passive and integrated all-optical nonlinear activation method—implemented solely in nanophotonic waveguides without bias signal, auxiliary power input, or additional material integration, and offering ultrafast response for high computational bandwidth together with light-speed, low-latency processing—represents a highly desirable pathway toward scalable, high-throughput PIC-based ONN architectures.

In this work, we experimentally demonstrate, a passive chip-scale all-optical nonlinear activation method exhibiting a sigmoid-like response, completely free from any external control or driving energy. 
The sigmoid function, a foundational nonlinearity in modern AI models \cite{4}, is realized through an efficient $\chi^{(2)}$ process in a periodically poled lithium niobate (PPLN) nanophotonic waveguide, achieving a high second-harmonic generation (SHG) conversion efficiency of approximately 80\%, approaching state-of-the-art reported values \cite{78}. 
This enables the all-optical nonlinearity essential for learning complex mappings in ONNs, with femtosecond-scale, wavelength-selective activation that supports computational rates exceeding 100 GHz—compatible with those state-of-the-art optical linear operations and allowing further gains through recent multiplexing techniques \cite{dong2024nre}.
To realise the corresponding neuron linear operation, we develop a silicon PIC tailored for arbitrary matrix–vector multiplications using MZI networks. By optically cascading a nonlinear PPLN chip with a linear MZI chip, we construct a photonic neuron capable of neural computing completely in the optical domain.
We demonstrate all-optical neuron-level inference within a single-hidden-layer network architecture, where the network operation is realized through time-multiplexed reuse of the optical neuron, and verify its robust ability to learn and represent nonlinear decision boundaries across canonical machine-learning benchmark tasks.
Then, we further investigate the broader applicability of $\chi^{(2)}$-enabled photonic nonlinearities in multi-layer neural network settings. Using the experimentally characterised PPLN activation response together with linear products computed by the silicon MZI-based PIC, we evaluate deeper SHG-activated ONNs on real-world machine-learning tasks. In an airfoil self-noise regression benchmark and a medical image classification task, the resulting ONNs achieve performance comparable to conventional electronic implementations, highlighting the potential of $\chi^{(2)}$ nonlinearities for scalable optical learning systems.
Overall, this work marks an important step toward fully integrated, all-optical deep neural networks by bridging highly efficient nanophotonic parametric nonlinear processes with rapidly evolving field of PIC-based optical computing.

\begin{figure}[!t]
\centering
\includegraphics[width=1\linewidth]{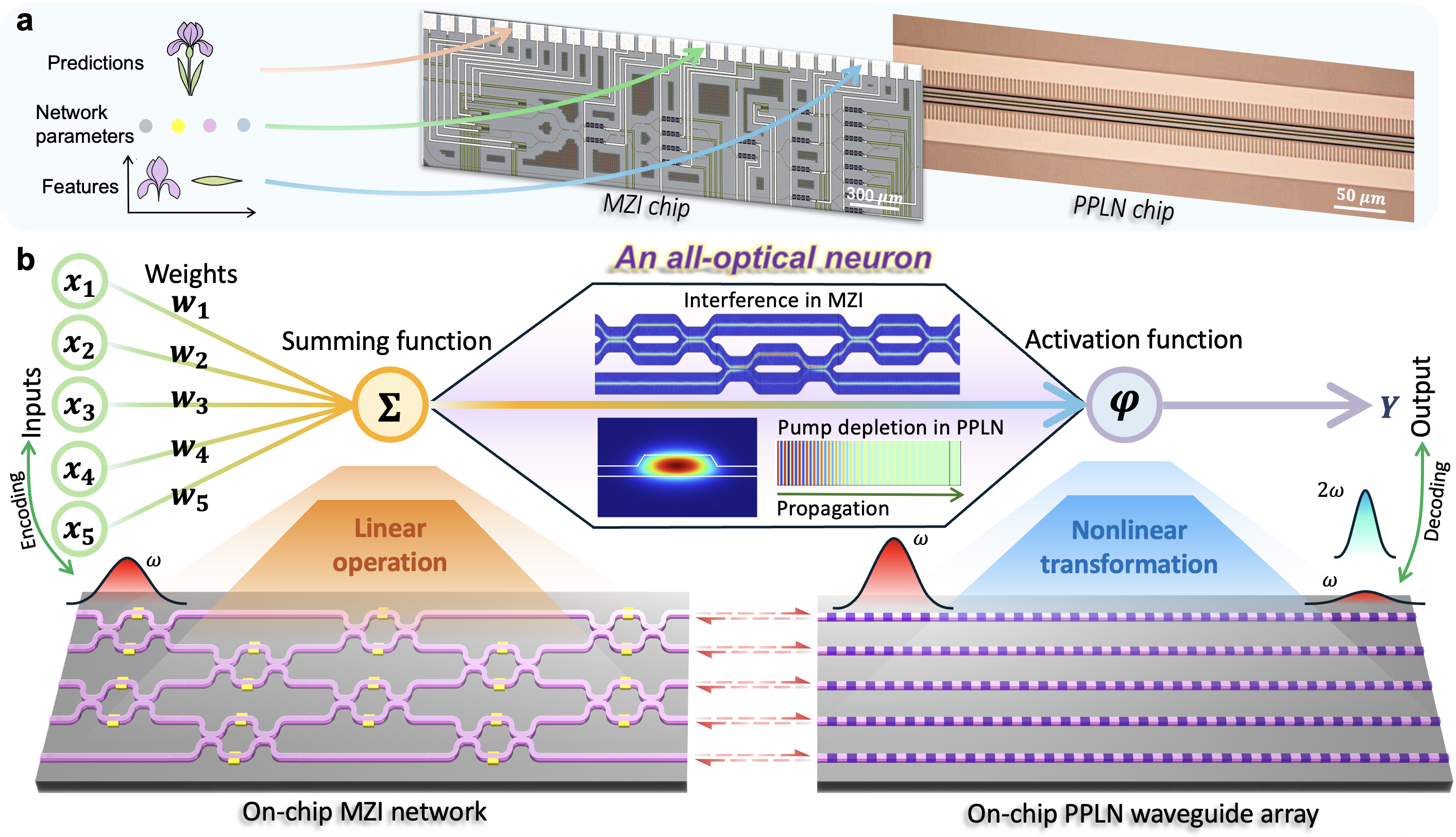}
\caption{\textbf{All-optical neuron architecture employing pump-depleted SHG in PPLN nanophotonic waveguides to provide on-chip nonlinear activation with MZI-based linear processing.} 
\textbf{a} Microscope images of the MZI and PPLN chips. Input features and network parameters are programmed into the phase shifters through electrical pads, and intermediate predictions are extracted from the on-chip optical output ports.
\textbf{b} Neuron inputs are encoded in the amplitudes and phases of light at the fundamental harmonic frequency $\omega$. These coherent optical signals propagate iteratively through the MZI network and the PPLN array, where dot products are executed via optical interference, and nonlinear activation is realized through efficient, passive pump-depleted SHG enabled by quasi-phase matching and strong $\chi^{(2)}$ parametric interaction from tight optical confinement in the TFLN waveguide. The neuron's output is extracted via homodyne detection at the same frequency.}
\label{fig_1}
\end{figure}

\begin{figure}[!ht]
\centering
\includegraphics[width=1\linewidth]{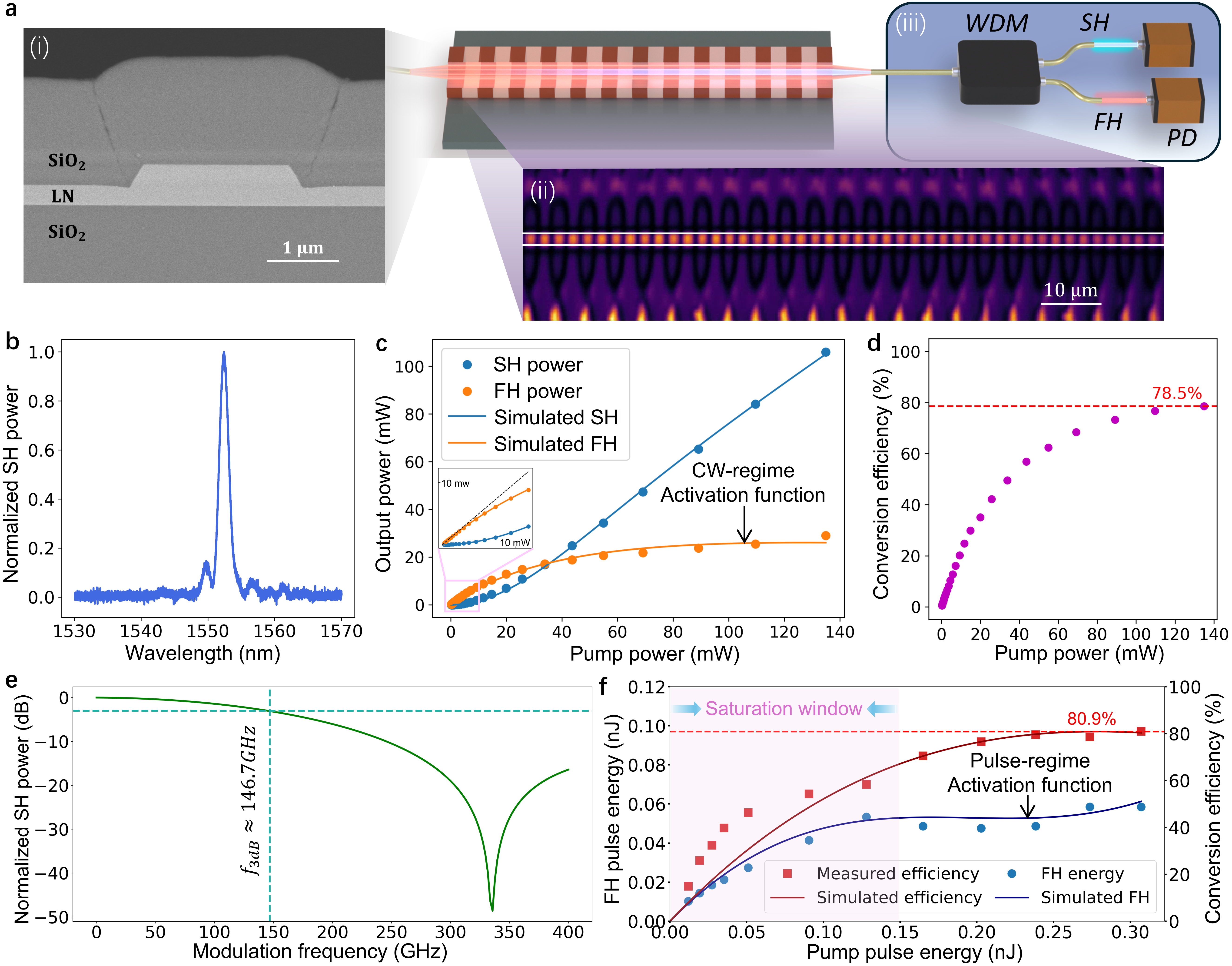}
\caption{\textbf{Pump-depleted SHG process in a PPLN nanophotonic waveguide for all-optical nonlinear neuron activation.} 
\textbf{a} PPLN device fabrication and characterization. (i) Cross-sectional SEM image. (ii) Top-view laser-scanning SHG microscope imaging. (iii) Measurement setup: output light is wavelength demultiplexed by a WDM into FH and SH channels and detected by photodiodes (PD).
\textbf{b} SHG spectrum of the fabricated PPLN waveguide, showing a phase-matching wavelength at 1552 nm.
\textbf{c} Measured CW-pumped nonlinear activation (dots) of the FH and SH light. Solid curves represent scaled Runge–Kutta simulation, revealing a saturable, sigmoid-like nonlinearity for the FH light with a nonlinear onset threshold of approximately 2 mW and a saturation window emerging below 10 mW (inset). 
\textbf{d} Calculated absolute conversion efficiency from the FH to SH as a function of CW pump power. 
\textbf{e} Numerical SH power transfer function versus pump modulation frequency, with the 3-dB activation bandwidth set by group-velocity mismatch. 
\textbf{f} Measured pulsed-pumped nonlinear activation of the FH light and corresponding conversion efficiency, showing a  nonlinear onset threshold near 0.02 nJ and a sigmoidal saturation window below 0.15 nJ.}
\label{fig_2}
\end{figure}

\begin{figure}[!t]
\centering
\includegraphics[width=1\linewidth]{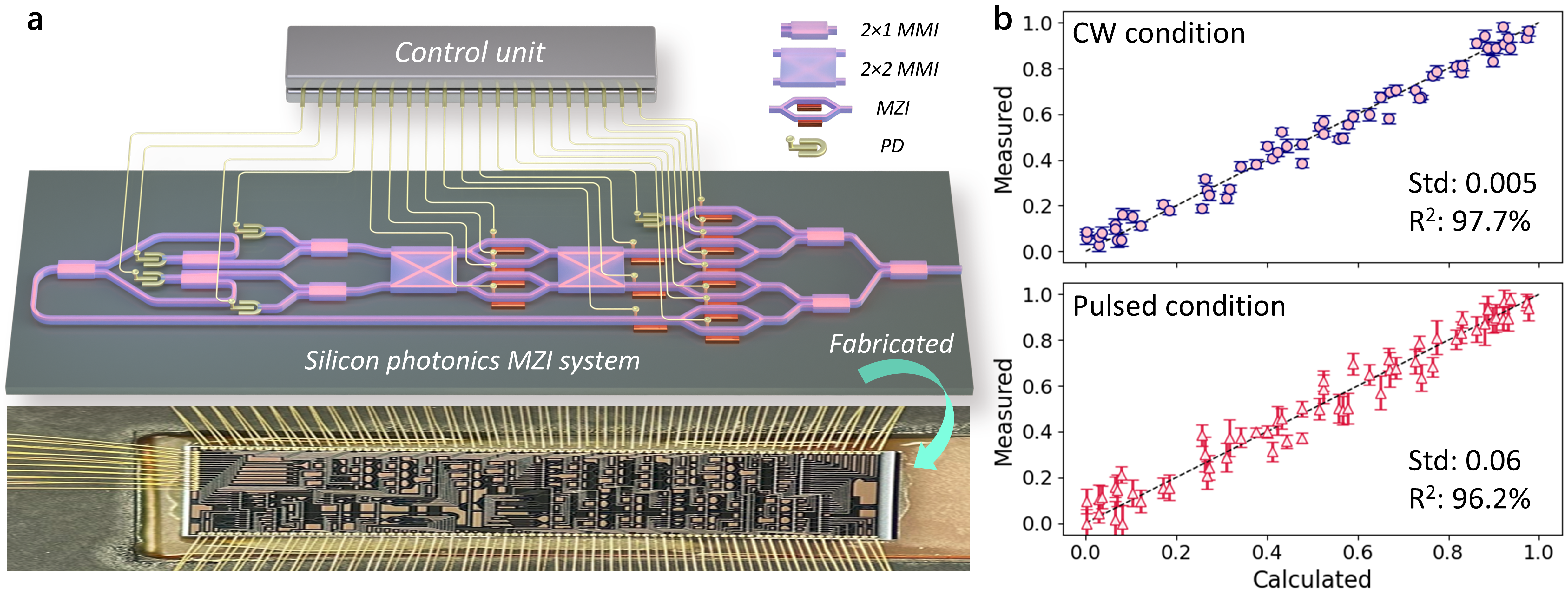}
\caption{\textbf{Design and characterization of the MZI system on a PIC for the linear operation required by the proposed optical neuron.}
\textbf{a} Schematic and microscopic image of the integrated silicon photonics-based MZI system, comprising on-chip multimode interferometers (MMIs), thermo-optic phase shifters, PDs, and wire bonding for electrical control. The system supports arbitrary $2\times2$ matrix–vector multiplication.
\textbf{b} Experimental characterization of the PIC-based optical matrix multiplication using randomly generated input matrices to evaluate linear computation fidelity. The measured outputs yield a standard deviation of 0.005 and an $R^2$ of 97.7\% under CW operation, and 0.06 and 96.2\%, respectively, under pulsed operation. These metrics confirm that the system achieves the fidelity necessary for the linear computations performed by the optical neuron.}
\label{fig_3}
\end{figure}

\subsection*{Pump-depleted $\chi^{(2)}$-based all-optical nonlinear neuron activation}

The core of our approach is the utilisation of the pump-depleted SHG process as all-optical nonlinear activation functions. 
Due to energy conservation and phase-matching conditions, the fundamental-harmonic (FH) wave undergoes efficient energy transfer into the second-harmonic (SH) wave with increasing pump power, resulting in pronounced pump depletion and a saturating FH response.
This interaction produces a sigmoid-like nonlinear transfer function (Supplementary Fig. S2), well-suited for neural network activation.

The activation is implemented on a thin-film lithium niobate (TFLN) platform, which offers high second-order nonlinear susceptibility that supports efficient $\chi^{(2)}$ interactions for wavelength conversion \cite{64,shi2024efficient,shi2025squeezed}. Its ferroelectric property makes possible electrical poling for domain inversion, allowing quasi-phase-matched PPLN nanophotonic waveguides for efficient SHG and substantial pump depletion \cite{wang2018ultrahigh}.
By encoding computational information into the FH wave, the proposed nonlinear activation achieves a femtosecond-scale response originating from the near-instantaneous nature of dipole polarization\cite{76}. As a result, it requires no external tuning or additional energy input, making the activation inherently all-optical, fully passive, and low-latency, while remaining fully compatible with PICs,

Figure \ref{fig_1} shows an all-optical neuron architecture, in which linear operations in MZI network and PPLN-based nonlinear activations alternate across layers, together forming the basis for optical neural inference. 
For clarity, only a single representative “linear-then-nonlinear” stage is shown as a canonical neuron. 
We note that the experimental system presented later adopts the inverse ordering where the PPLN activation precedes the MZI-based linear computation; however, this does not alter the functional equivalence of the neuron or the generality of the computational model.
Typically, the system proceeds in three computational steps. 
(1) Following Clement’s optical computing architecture \cite{20}, the input data vector 
$x = \{x_1, x_2, \dots, x_n\}$
is encoded into the FH at frequency $\omega$, with the normalized data magnitude and sign represented by the amplitude \( A_n \) and phase \( \phi_n \) of the transverse electric field component  $E_n(\omega) = A_n e^{i(\omega t + \phi_n)}$. 
(2) The weight matrix is programmed via voltages applied to phase shifters within the MZI system. As the optical signals propagate through the network, matrix–vector multiplication is intrinsically performed through passive interference.
(3)  The output signals from the MZI system are utilized as the pump for the PPLN waveguides, where they undergo a pump-depleted SHG process. This induces a strong nonlinear transformation of the FH wave carrying the computational information, as described by Eq.~(\ref{eq1}), thereby realizing passive all-optical activation.
\begin{equation}
    E(\omega, L) = E(\omega, 0) \frac{1}{\sqrt{\cosh^2 (\Gamma L) + (\Delta k / 2\Gamma)^2}} e^{-\alpha_{\omega} L}
\label{eq1}
\end{equation}

\noindent where \( E(\omega, 0) \) denotes the initial FH amplitude encoding the weighted sum of the neuron, and \( E(\omega, L) \) represents the FH amplitude after the nonlinear interaction over a propagation length of $L$. The nonlinear coupling coefficient under quasi-phase matching conditions is given by $\Gamma = \frac{4\omega d_{\text{eff}} E(\omega, 0)}{\pi{\sqrt{n_{\omega}n_{2\omega}}c}}$, where \( c \) is the speed of light in vacuum, \( d_{\text{eff}} \) is the effective second-order nonlinear coefficient, and \( n_{\omega} \) and \( n_{2\omega} \) are the effective indices of the FH and SH modes, respectively \cite{76}. In an ideal case of negligible phase mismatch ($\Delta k = 0$) and loss ($\alpha_{\omega} = 0$), the nonlinear activation function for the FH wave simplifies to $E(\omega, L) = E(\omega, 0) \text{sech} (\Gamma L)$, yielding a saturating sigmoid-like nonlinear transformation over a defined propagation length.
This approach is inherently compatible with coherent photonic computing architectures that encode data in field amplitude and phase.

\section*{Results}

\subsection*{Fabrication and characterization of nonlinear and linear PICs}

A 11-mm-long PPLN nanophotonic waveguide is fabricated to achieve the nonlinear transformation required for the activation function in the proposed optical neuron (see fabrication details in Methods).
Figure \ref{fig_2}\textbf{a}(i) shows a scanning electron microscopy image of the waveguide cross-section, revealing the ridge geometry that ensures strong optical confinement for efficient nonlinear interaction. Figure \ref{fig_2}\textbf{a}(ii) presents a top-view confocal SHG imaging, confirming the uniform periodic domain inversion, allowing efficient quasi-phase matching along the waveguide.
The experimental setup for characterizing the SHG response is illustrated in Fig. \ref{fig_2}\textbf{a}(iii), where the generated SH and residual FH signals are separated by a wavelength-division multiplexer (WDM) and detected by photodetectors.
We measured the SHG spectrum of the PPLN waveguide, with the normalized SHG intensity shown as a function of the pump wavelength in Fig. \ref{fig_2}\textbf{b}, revealing a phase-matching FH wavelength of 1552 nm within the telecom C band.
With continuous-wave (CW) pumping, the output SH power and FH power are measured as functions of the on-chip pump power (Fig. \ref{fig_2}\textbf{c}).   
Experimental data points are fitted using a curve derived from fourth-order Runge–Kutta simulations of the coupled nonlinear wave equations governing pump-depleted SHG described by Eq.~(\ref{eq1}), showing good agreement between the experimental results and the theoretical model.
In the low-pump-power linear region, the device shows a normalized on-chip SHG conversion efficiency as high as 1986\% $\pm$ 26\% W$^{-1}$cm$^{-2}$, where the efficiency is calculated after excluding all fiber–chip coupling losses and referenced solely to the power inside the waveguide.
The conversion efficiency starts to saturate at a pump power of several milliwatts, and reaches 78.5\% (absolute) at 135 mW (Fig. \ref{fig_2}\textbf{d}).  
Unlike phase-change or photorefractive nonlinearities that exhibit a distinct physical activation threshold, the proposed $\chi^{(2)}$-based method is physically thresholdless, with the observable onset of nonlinear response appearing at pump powers of approximately 2 mW and a saturation curvature emerging below 10 mW (Fig.~\ref{fig_2}\textbf{c} inset).
As a result, the output power of the FH wave approximates a sigmoid-like nonlinearity within the measured power range, demonstrating the effectiveness of pump-depleted SHG in the PPLN waveguide as an all-optical nonlinear activation for the proposed optical neuron.  
The SHG transfer function (Fig.~\ref{fig_2}\textbf{e}), obtained from a frequency-domain model including group-velocity mismatch (GVM) and quasi-phase matching, predicts a 3-dB activation bandwidth of $\sim$146~GHz (see Methods), underscoring the intrinsically high-speed characteristic of PPLN-based neuron activation. 
To evaluate the intrinsic energy efficiency, we further characterize the FH nonlinear activation function under pulsed-pumping condition. Using a femtosecond laser (78~fs, 75~MHz) followed by narrowband filtering (0.92~nm FWHM bandwidth at 1552~nm) and amplification with an erbium-doped fiber amplifier, we achieved a conversion efficiency of 80.9\% at $\sim$0.3~nJ on-chip pulse energy (Fig.~\ref{fig_2}\textbf{f}). 
Pulsed excitation, which is compatible with coherent optical computing architectures \cite{yu2025wdm}, produces a more efficient nonlinear activation response and exhibits an inflection point corresponding to a curvature flip behavior, arising under near-ideal phase-matching conditions (Supplementary Note~I). 
This result suggests that operating in the pulsed regime could largely reduce the activation power, and highlights the potential for more than an order-of-magnitude improvement in energy efficiency.
For FH loss characterization, at pump levels approximately twice the nonlinear-onset threshold representative of the activation’s typical operating regime, the activation loss due to SHG conversion is measured to be 1.88 dB at 20 mW (CW) and 2.2 dB at 0.04 nJ (pulsed). These values are comparable to or lower than those reported for other advanced all-optical activation mechanisms \cite{107,44}, and they fall well within the cascaded loss budget required for deep ONN implementations (Supplementary Note III). Moreover, benefiting from the efficient $\chi^{(2)}$ process in PPLN, simulation results explicitly show that operation within this onset-adjacent regime already provides functionally meaningful nonlinear activation for network learning (Supplementary Fig. S7).

Figure \ref{fig_3}\textbf{a} illustrates the MZI network used for linear operations, on a 220 nm silicon-on-insulator (SOI) platform by Advanced Micro Foundry (Singapore). 
Thermo-optic phase shifters and on-chip germanium photodetectors are integrated to enable reconfigurable phase control and high-responsivity readout in the near-infrared, respectively.
The input signal is equally split into four channels using a series of 1×2 multimode interferometers, with two channels dedicated to subsequent computations and two reserved for detection and calibration. 
The design overall implements a two-dimensional arbitrary matrix–vector multiplier based on the Clements architecture \cite{20}, and serves as the linear computing module in the proposed optical neuron.
To characterize the MZI system's performance, we conducted 300 measurements by supplying 60 sets of randomly generated $2 \times 2$ matrices (with elements normalized between 0 and 1), each repeated five times, and compared the measured outputs to the ground truth. 
The system achieved a standard deviation of 0.005 and an $R^2$ of 97.7\% under CW operation, and a standard deviation of 0.06 and an $R^2$ of 96.2\% under pulsed operation, demonstrating high accuracy in performing analog matrix–vector multiplications (Fig.~\ref{fig_3}\textbf{b}). 
These results confirm the suitability of the developed MZI system as a robust weighted-sum module in the proposed all-optical neuron architecture.

\begin{figure}[!t]
\centering
\includegraphics[width=\linewidth]{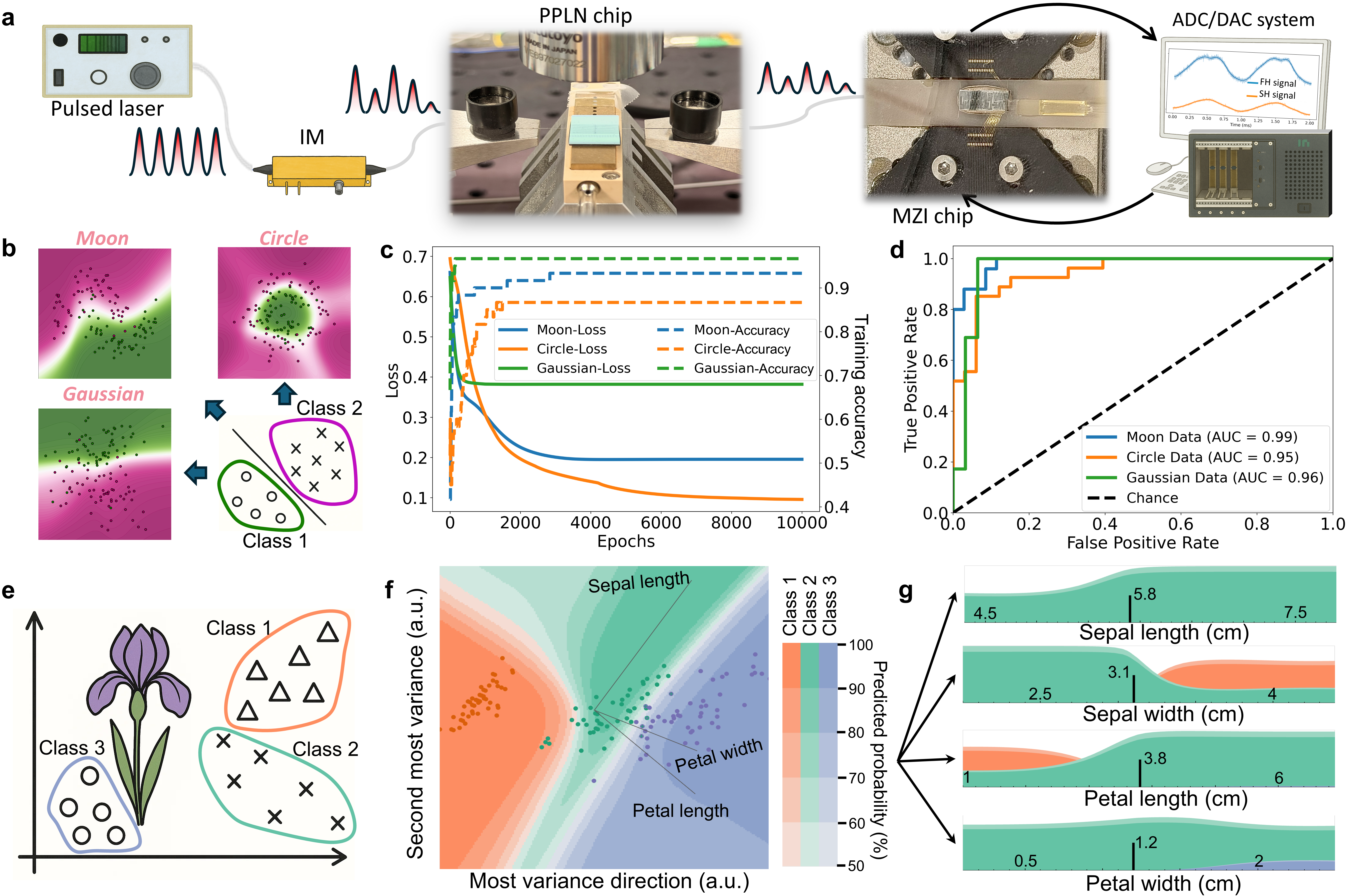}
\caption{\textbf{Demonstration of an all-optical neuron by optically cascading the PPLN and MZI chips for single-hidden-layer ONN learning.} 
\textbf{a} Pulsed pump at phase-matched FH wavelength is launched to an intensity modulator (IM), where the weighted neuron input values are encoded using an arbitrary waveform generator operating at 1 kHz. The modulated light is subsequently coupled into a PPLN chip to realize $\chi^{(2)}$-based nonlinear activation. The nonlinearly transformed pulse is then routed to a silicon MZI chip, which performs the weighted multiplication, and the on-chip inference output is detected at the on-chip photodetection readout.
    \textbf{b} Decision boundaries generated by the SHG-activated ONN for three binary classification datasets: Moon, Circle, and Gaussian. Pink and green dots represent two separable sets. 
\textbf{c} Training loss and accuracy curves over 10000 epochs in the differentiable physics-aware digital twin of the ONN. 
\textbf{d} ROC curves and AUC scores of the ONN inference demonstrate high classification performance: AUC = 0.99 (Moon), 0.95 (Circle), and 0.96 (Gaussian), significantly outperforming the chance level (dashed line). 
\textbf{e} Schematic diagram of Iris flower dataset consisting of three species. 
\textbf{f} Two-dimensional map of ONN’s decision surface around training data points, with axes representing directions in feature space based on principal component analysis. 
\textbf{g} Partial dependence shows predicted probability for a class change along varied feature value. Black vertical line represents the current feature value of the selected instance.}
\label{fig_4}
\end{figure}

\subsection*{$\chi^{(2)}$-based nonlinear learning with an all-optical neuron}
To evaluate the nonlinear learning capability of the proposed $\chi^{(2)}$-based ONN, we implement a proof-of-concept all-optical neuron inference experiment (Fig.~\ref{fig_4}\textbf{a}). 
In this setup, the neuron inputs are encoded onto pulsed light ($\sim$0.25 nJ) using a fiber-based intensity modulator and then coupled into the PPLN chip for nonlinear activation. The resulting optical signal is subsequently coupled via single-mode fiber into the silicon MZI network, which performs the neuron’s weighted-sum operation.
The system thus follows an activation-first, weighting-later architecture, a neuron-level computational ordering that has been shown to support efficient learning and is similar to recently proposed Kolmogorov–Arnold networks (KANs) \cite{reinhardt2025sinekan}. This nonlinear–linear optical sequence forms a single all-optical neuron module. By time-multiplexing this neuron, a single-hidden-layer ONN can be constructed, enabling direct evaluation of SHG-activated optical inference without any electro–optical conversion within the neuron (see Methods).

A single hidden layer with 32 neurons is used as a testbed to quantify the learning capability of the SHG-activated ONN on multivariate datasets.
The network is trained off-chip using a digital twin of the optical system (Supplementary Note II), where both the inner coefficients and outer weights of the ONN are optimized through gradient-based learning incorporating the experimentally calibrated SHG and MZI transfer functions.
This yields network parameters that are already matched to the physical hardware.
After training, the learned linear weights are programmed onto the silicon MZI network via thermo-optic phase shifters (see Methods), such that the chip performs purely optical inference during deployment.
Although the intrinsic SHG response in PPLN supports bandwidths well above 100 GHz, the present proof-of-concept system operates at an effective data rate of 1$\sim$ kHz, limited by the slow response of the MZI thermo-optic modulators and associated electronic systems. This constraint is not fundamental to the optical activation itself and can be lifted to gigahertz regime by migrating to a fully integrated TFLN platform with high-speed electro-optic (EO) modulation (see Discussion).
To realize the 32-neuron hidden layer using a limited number of physical interferometers, the weighted summation is decomposed into element-wise multiplications and executed in a time-multiplexed manner across the optically connected chips, enabling fast completion of a full 32-neuron single-layer ONN inference within approximately 1 s (inclusive of all electronic control and conversion).

We first assess the model on three canonical binary classification benchmarks—Moon, Circle, and Gaussian datasets—from the Scikit-learn library \cite{90}, each containing 120 samples evenly split for training and testing. Figure. \ref{fig_4}\textbf{b} shows the raw data distributions with learned decision boundaries, where the $\chi^{(2)}$-driven ONN successfully adapts to varying geometric structures. 
The training loss decreases steadily while the test accuracy continues to rise, indicating reliable convergence and minimal overfitting of the SHG-based activation (Fig. \ref{fig_4}\textbf{c}). The receiver operating characteristic (ROC) curves (Fig. \ref{fig_4}\textbf{d}) further confirm strong discriminative capability, yielding area under the curves (AUCs) of 0.99 (Moon), 0.95 (Circle), and 0.96 (Gaussian). Correspondingly, the final test accuracies reach 0.91, 0.89, and 0.97, demonstrating that the SHG-driven all-optical activation reliably forms effective decision boundaries across diverse binary classification tasks. These results are also comparable to state-of-the-art SOI-based \cite{pai2023MZI_circle} and TFLN-based \cite{zheng2024MZI_circle} ONNs that rely on electronic nonlinearities with in-situ training, further highlighting that the proposed all-optical activation scheme can be integrated and scaled alongside the recent advanced ONN platforms.

We use the Iris dataset from Scikit-learn (Fig. \ref{fig_4}\textbf{e}) to evaluate the SHG-activated ONN's ability to generate decision surfaces in higher-dimensional feature space.
Three features are activated to reduce visual complexity while preserving model fidelity (Fig. \ref{fig_4}\textbf{c}). The resulting decision map shows well-separated regions for each class with distinct boundaries, demonstrating clear class confidence zones.
The partial dependence (Fig. \ref{fig_4}\textbf{g}) reveal smooth and monotonic class transitions for each active feature. Probability shifts near decision thresholds reflect the proposed activation’s high sensitivity, enabling precise class separation with minimal variation in input features. Together, these results confirm the $\chi^{(2)}$-driven ONN's capability to construct expressive, localized decision boundaries, leading to a final classification accuracy of 96.7\% on this multi-class task.

\begin{figure}[!t]
\centering
\includegraphics[width=1\linewidth]{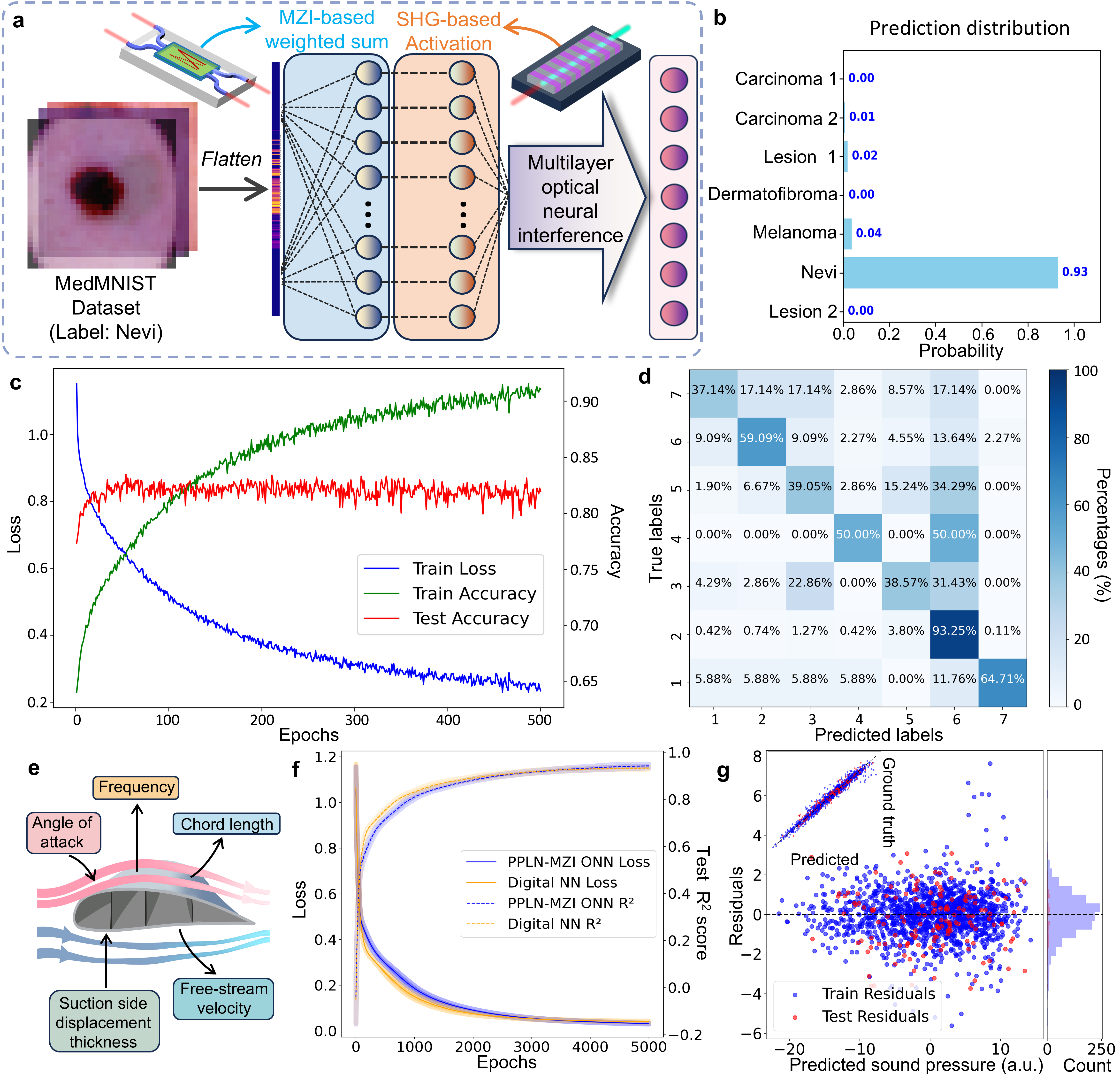}
\caption{
    \textbf{Multi-layer ONNs with measured $\chi^{(2)}$-based activations in hybrid electro-optical neural computing.}
    \textbf{a} Inference pipeline of the proposed optical neurons on the MedMNIST dataset. The input image is first flattened into a vector and encoded into the amplitudes and phases of optical signals. These signals undergo a linear weighted-sum operation via MZI-based interference, producing hidden-layer representations. The outputs are then nonlinearly activated using the fitted FH response curve derived from experimentally measured pump-depleted SHG data. After passing through multiple layers of MZI–SHG-based transformations, the final output is fed into a softmax layer to generate the prediction distribution.
    \textbf{b} Example prediction probability distribution for a nevi class generated by the ONN.
    \textbf{c} Training loss, accuracies, and test accuracy over epochs on the DermaMNIST-C dataset. 
    \textbf{d} The confusion matrix of the MLP model trained on the  dataset.
    \textbf{e} A schematic representation of the Airfoil Self-Noise dataset with five aerodynamic features.
    \textbf{f} Training loss and test $R^2$ score over 5,000 epochs, demonstrating the ONN's stable convergence. 
    \textbf{g} Residuals plotted against predicted values, with an accompanying histogram showing their distribution centered around zero. The inset displays predicted versus true sound pressure levels, demonstrating strong agreement along the ideal diagonal, indicating accurate model fitting.
    }
\label{fig_5}
\end{figure}

\subsection*{Classification on dermatological images by ONNs}

AI systems are increasingly effective in practical classification tasks. To evaluate the potential capability of our proposed $\chi^{(2)}$-based optical neuron in real-world scenarios, we employ the experimentally measured CW-pumped SHG-induced FH response curve as a surrogate activation function. This activation function is then implemented in an electronic backend, which complements the MZI mesh that performs the required linear operations, to validate multi-layer ONN learning and assess the stability of the PPLN-driven activation method.
We demonstrate the system on the DermaMNIST-C dataset \cite{75}, a 28×28-pixel dermatology subset of MedMNIST \cite{74} containing seven disease categories that together account for approximately 95\% of clinically observed skin lesions.
We implement a multilayer perceptron (MLP) as the baseline model, a foundational building block to model high-dimensional feature representations in modern AI systems. Each image is flattened into a 784-dimensional input vector and processed through a SHG-activated MLP with three hidden layers (256, 128, 64 neurons) (Fig. \ref{fig_5}\textbf{a}). The final hidden layer output is passed through a digital softmax layer to generate class probability distributions (Fig. \ref{fig_5}\textbf{b}).

The training dynamics are illustrated in Fig. \ref{fig_5}\textbf{c}, showing the evolution of loss and accuracy over epochs. During early training, the test accuracy briefly exceeds the training accuracy, indicating an effect commonly observed in small or imbalanced real-world datasets. 
As training progresses, the training accuracy steadily increases, whereas the test accuracy saturates early once the model has captured the dominant generalizable features, such as those associated with the Nevi class. This trend is consistent with previous analyses on DermaMNIST, where class imbalance and limited test-set diversity can lead to elevated early test accuracy \cite{75}. The final test accuracy reaches 82.64\%, closely matching the reported performance of a digital ResNet-18 (82.50\%) \cite{75}, demonstrating that the $\chi^{(2)}$-based ONN generalizes effectively and captures clinically meaningful features in this complex dermatological classification task.
The confusion matrix (Fig. \ref{fig_5}\textbf{d}) shows strong performance on Nevi (93.35\%), the largest class, and reasonable accuracy on Lesion 2 (64.71\%) and Carcinoma 2 (61.36\%), indicating the model’s ability to learn subtle diagnostic patterns.
Overall, these results validate the effectiveness of the proposed $\chi^{(2)}$-based nonlinear activation, jointly operating with MZI-based linear transformations, in enabling an ONN that attains performance comparable to digital implementations on high-dimensional clinical-image datasets.

While the present demonstration uses a hybrid ONN with experimentally measured PPLN activations, fully all-optical deployment of such large-input-dimension tasks will benefit from multiplexing strategies, such as time or wavelength multiplexing, to increase throughput and reduce power consumption. For example, in a four-channel optical core operating at 20\,mW per channel (sufficient for both CW and pulsed activation), the total on-chip power is $\sim$80\,mW, scaling linearly with channel count (e.g., $\sim$2\,W for 100 channels). This $\sim$2 W total optical power level is comparable to that of state-of-the-art optical neural processors\cite{hua2025pace}, indicating that the proposed activation scheme remains within a practical power envelope for large-scale photonic ONN implementations.

\subsection*{Regression on the airfoil self-noise dataset by ONNs}

To further examine the versatility of our $\chi^{(2)}$-based ONN beyond classification tasks, we apply the same optical computing architecture, MZI-based linear transformations with the measured $\chi^{(2)}$ nonlinear activation, to a regression task. Specifically, we evaluate its compatibility with continuous-valued prediction using the NASA Airfoil Self-Noise dataset \cite{62}. For this task, we implement a compact two-hidden-layer MLP (64, 64 neurons). The dataset involves predicting sound pressure levels from five aerodynamic features (Fig. \ref{fig_5}\textbf{e}), offering a distinct yet practically relevant benchmark for assessing the ONN’s generalization across heterogeneous real-world tasks.

Figure \ref{fig_5}\textbf{f} presents the training and evaluation process, showing the training loss alongside the evolving $R^2$ score on the test set. The $\chi^{(2)}$-based hybrid ONN exhibits stable convergence and ultimately achieves an approximate $R^2$ of 0.94, indicating strong predictive capability. By approaching the performance of its digital counterpart with same network architecture and electronic sigmoid activation, this result highlights the practical viability of our optical neuron architecture for complex regression tasks.
Figure \ref{fig_5}\textbf{g} compares residuals with predicted values, and the accompanying histogram shows a narrow, symmetric distribution centered near zero, suggesting minimal systematic bias. The inset displays predicted versus true noise levels for both training and test sets, showing close alignment with the diagonal and further validating regression accuracy.
Together, these results demonstrate that the SHG-activated optical neuron, combined with MZI-based linear operations, can effectively capture nonlinear relationships in real-world data.

\section*{Discussion}
We present a passive all-optical nonlinear activation method based on strong experimentally demonstrated $\chi^{(2)}$ nonlinearity in a PPLN nanophotonic waveguide, achieving high absolute SHG conversion efficiencies of $\sim$80\%.
The realised parametric process induces an ultrafast, sigmoid-like nonlinear response through depletion of the FH light carrying computational information, without requiring additional signals or energy input.
Leveraging the maturity of silicon photonics, we demonstrate an all-optical neuron architecture that concatenates the passive PPLN-driven nonlinear activation with a silicon MZI network for optical linear computation.
Experimental results show that the resulting ONN can capture complex nonlinear decision boundaries in synthetic binary-classification benchmarks and reaches 96.7\% accuracy on the Iris dataset.
Moreover, by integrating the experimentally measured activation curve into an electronically trained backend and still using the MZI network for optical linear transformations, SHG-activated multilayer ONNs achieve accuracies comparable to state-of-the-art electronic models on real-world machine-learning tasks, including medical-image classification and airfoil-noise regression. These findings confirm the robustness and feasibility of the proposed activation scheme and underscore its potential as a foundation for scalable, all-optical ONNs.

From a machine learning perspective, the sigmoid function remains a foundational activation due to its boundedness, differentiability, and biological plausibility \cite{57}. However, efficient all-optical realizations of sigmoid-shaped responses within PIC-based ONNs are scarce. In electronic systems, computing a single sigmoid-shaped function typically incurs a delay on the order of tens of nanoseconds \cite{55}, generally contributing minimally to the overall inference time dominated by dense linear operations.
In contrast, ONNs can perform all-optical linear transformations at computational frequencies exceeding 100 GHz\cite{hu2025NPR,11}, offering up to 100-fold faster processing than conventional electronics.
To fully harness the computational potential of photonics, it is essential to realize nonlinear operations that match this speed and exhibit an ultrafast response, thereby enabling full exploitation of the vast bandwidth inherent to light.
To facilitate a comprehensive comparison, we summarize existing all-optical nonlinear activation mechanisms with sigmoid-like characteristics in Table \ref{tab1}, focusing on six key metrics. Passivity is critical for energy efficiency, reduced control complexity, and architectural simplicity. Response time reflects the  material's intrinsic speed in reacting to light, governing how quickly individual operations can be executed. Computational bandwidth represents the upper bound of sustained processing capacity, determining the system's maximum achievable throughput. 
Time-of-flight sets the minimal latency of a single activation event.
Wavelength selectivity provides opportunities for wavelength-division multiplexing to enhance parallelism.
Waveguide-only design shows whether the nonlinearity can be achieved in pure waveguides without extra material integration, supporting fabrication simplicity and deep ONNs on a unified photonic chip.

Most previously reported optical nonlinear activation methods rely on active modulation schemes, with response speeds fundamentally constrained by carrier dynamics or electronic control circuitry. As a result, they typically exhibit response times on the order of tens of picoseconds, limiting their computational frequencies to below several tens of gigahertz. 
In contrast, our approach leverages a fully passive $\chi^{(2)}$ parametric process with an intrinsic nonlinear optical response in the femtosecond regime, where the $<100$-fs value listed in Table~1 represents the upper-limit estimate of this almost instantaneous response governed solely by electronic polarization dynamics \cite{76,schumacher2020pnas}.
While the current device bandwidth could be limited to a few hundred gigahertz by GVM and quasi-phase-matching conditions, it is already comparable to that of state-of-the-art EO modulators and does not represent a fundamental limit. The underlying $\chi^2$ process is intrinsically near-instantaneous, implying a much broader bandwidth ceiling for FH nonlinear activation. Looking ahead, dispersion/GVM engineering, adaptive poling strategies, and systematic loss reduction could substantially extend the effective computational bandwidth. 
Similarly, the relatively long time-of-flight delay of the current 11-mm PPLN waveguide does not present a computational bottleneck, since optical computing derives its benefits primarily from high throughput rather than propagation speed \cite{10}.  
Moreover, our proposed activation is implemented using a compact straight waveguide geometry without requiring additional epitaxy \cite{41,44,52}, facilitating fabrication simplicity, minimal optical footprint, and excellent cascadability, all of which are critical for future multilayer ONN scalability.

From a footprint perspective, the waveguide-only architecture affords substantial geometrical flexibility: the 11-mm device can be folded into compact layouts through multi-segment routing with adiabatic bends, reducing the lateral footprint to the millimeter scale while preserving the micron-scale vertical dimension—similar to the densely cascaded EO modulator arrays commonly realised on TFLN platforms.
For a system-level energy--efficiency outlook, the proposed PPLN activation operated in the pulsed regime typically requires an energy of $\sim 0.25\,\mathrm{nJ}$ (corresponding to 18\,mW average power in our settings) and inherently supports operation at gigahertz level, enabling high-throughput optical computing. Treating each activation as a single operation and assuming a readily achievable $>\!20\,\mathrm{GHz}$ EO modulation rate on LN platform, the resulting computing efficiency exceeds $1\,\mathrm{TOPS/W}$, matching state-of-the-art linear optical computing systems \cite{hua2025pace}. 
The intrinsic wavelength selectivity of the SHG process also enables compatibility with wavelength-division multiplexing, offering the potential to further increase data throughput \cite{43}.
Taken together, these features position our approach as a promising pathway toward practical, high-speed, and scalable integrated all-optical neural networks.

The current all-optical neuron prototype demonstrates all-optical learning by cascading a pump-depleted SHG process in PPLN waveguides for nonlinear activation with a silicon-photonics MZI network for linear operations, establishing the feasibility of performing neural inference entirely in the optical domain.
Nevertheless, operating across two discrete chips imposes limitations similar to those encountered in electronic systems, including power inefficiencies, timing overhead, optical power mismatches, and additional coupling losses.
Future integration efforts may leverage heterogeneous silicon–lithium niobate platforms, where linear MZI networks and nonlinear PPLN sections are co-fabricated through hybrid bonding or optimized taper coupling \cite{63}. Such hybrid architectures can reduce interfacial optical losses and improve power compatibility between the two subsystems.
A more compelling and increasingly practical direction is to migrate both the linear and nonlinear devices onto a monolithic TFLN platform, which inherently supports low-loss waveguides (typically $\sim$0.1 dB/cm), high-speed EO modulation (approaching $\sim100$ GHZ), and quasi-phase-matched nonlinear interactions within a single material system \cite{64}. This single-platform approach offers minimal optical loss, simplified fabrication, and multifunctional on-chip implementation of all required components, making it highly suitable for scalable ONN architectures.
Extensions such as multilayer integration and system-level acceleration represent natural next steps, and our approach is highly complementary to recent advances in TFLN PICs leveraging high-speed EO modulation \cite{song2025dac}, together pointing toward a broad and promising landscape where passive ultrafast activation function and scalable photonic processors can be co-developed.
While the proposed PPLN activation is primarily designed for the single-wavelength optical computing architectures that dominate current deep ONN demonstrations \cite{52,ashtiani2022deep,yan2025deep}, it can also be adapted to WDM-based computing schemes. In WDM-distributed architectures where each wavelength is routed into an individual wavelength-dependent channel~\cite{ou2025WDM}, the PPLN activation can be directly cascaded after each linear summation without spectral conflict. For clustered WDM channels, multi-wavelength activation could also be achieved as a potential extension by employing resonator-enhanced PPLN structures or by cascading multiple quasi-phase-matching periods or chirped-poling designs, albeit at the cost of increased device length.
Moreover, optimized pulse-pump schemes for data encoding and computation could reduce overall energy consumption by employing shorter, higher-peak-power optical pulses, thereby enhancing the efficiency of all-optical nonlinear activation.
In the future, PPLN microring resonators may also help reduce the required pump power and overall footprint by over an order of magnitude, owing to strong intracavity field enhancement and long effective interaction length enabled by the resonant geometry.

We note that practical deep ONNs do not require reproducing every digital operation exactly; instead, each optical layer must furnish a meaningful nonlinear transformation, with the overall network trained using emerging in-situ photonic optimization schemes \cite{momeni2025training}. Under this framework, maintaining stable and well-behaved per-layer nonlinearity is far more critical than preserving absolute optical power through many stages. Our quantitative FH-loss and onset-power analysis shows that the present $\chi^{(2)}$-based activator exhibits manageable loss and predictable operating conditions, indicating that multiple nonlinear layers can, in principle, be cascaded without intermediate optical amplification. This suggests that the approach is scalable toward multilayer ONN implementations.
This work represents a milestone experimental realization of a chip-based, passive all-optical nonlinear neuron enabled by $\chi^{(2)}$-driven activation functions with femtosecond response and a waveguide-only architecture, establishing the foundation for ultrafast  and scalable multilayer ONNs. 
Nonetheless, achieving a fully integrated all-optical neural network operating at gigahertz-scale speeds on a single chip remains a crucial next step toward next-generation neuromorphic photonic computing.

\begin{table}[!t]
\centering
\caption{Comparison of state-of-the-art sigmoid-shaped optical nonlinear activation functions. 
}
\begin{tabular}{cccccccc}
\toprule
& \makecell{Passivity} & \makecell{Response\\time} & \makecell{Computational\\bandwidth} & \makecell{Time-of-flight} & \makecell{Wavelength\\selectivity} & \makecell{Waveguide-only \\design} & \makecell{Integration\\scalability} 	 \\
\midrule
SOA \cite{41} & No & $\sim$100 ps & $\sim$10 GHz & $\sim$10 ps & No & No & Moderate \\
MRR \cite{42,38} & No & $\sim$400 ps & $\sim$2.5 GHz  & $\sim$10 ps & No & Yes & High \\
Ge-Si PD \cite{44} & Yes & 51 ps & $\sim$20 GHz & 0.11 ps & No & No & High \\
Modulator \cite{52} & No & $\sim$100 ps & $\sim$5 GHz & 105 ps & No & No & Moderate \\
SBS \cite{43}  & No & $\sim$ ns & $\sim$ sub-GHz & 500 ns & Yes & No & Early-stage \\
Polariton \cite{gan2025ep}  & No & $\sim$20 ps & $\sim$50 GHz & $<$ 1ps & Yes & No & Early-stage \\
SLM \cite{rahman2025Ozcan}  & Yes & $\sim$ ms
 & $\sim$kHz & $\sim$ 50ps & No & No & Early-stage \\
This work & Yes & $<$ 100fs \cite{76,schumacher2020pnas} & $>$100 GHz & 81 ps & Yes & Yes & High \\
\bottomrule
\label{tab1}
\end{tabular}
\end{table}

\section*{Methods}
\subsection*{PPLN nanophotonic waveguide fabrication}

We fabricate the PPLN nanophotonic waveguides on a 600 nm thick MgO-doped x-cut TFLN chip (NanoLN). 
Comb-like electrodes with a pitch of 4.75 \textmu m and duty cycle of 50\% are patterned using a combination of electron-beam (e-beam) lithography, followed by e-beam metal evaporation and lift-off. 
We apply a series of electrical pulses to periodically reverse the polarity for PPLN waveguides. 
A top-view laser-scanning SHG imaging of the PPLN waveguide is shown in Fig. \ref{fig_1}\textbf{a}(ii). 
After poling, the waveguides are patterned using e-beam lithography and Ar$^+$ etching by inductively coupled plasma reactive ion etching (ICP-RIE), with hydrogen silsesquioxane (HSQ) e-beam resist as the etching mask. 
Finally, the chip is covered with 1.65 \textmu m silicon oxide by plasma-enhanced chemical vapor deposition (PECVD). 
A cross-section SEM image is shown in Fig. \ref{fig_1}\textbf{a}(i).

\subsection*{PPLN computational bandwidth modeling}
To quantify the computational bandwidth of the pump-depleted SHG process in the quasi-phase-matched PPLN nanophotonic waveguide, we modeled the frequency response by evaluating the coherent build-up of the nonlinear polarization along the device length $L$ in the presence of GVM and residual phase detuning. The effective transfer function is defined as
\begin{equation}
H(\Omega) \;\propto\; \int_{0}^{L} \exp\!\left[i\,\Delta k(\Omega)\,z \;-\; \alpha z \right] \,\mathrm{d}z,
\end{equation}
where $\Delta k(\Omega)$ is the effective phase mismatch and $\alpha$ is the effective amplitude decay constant that accounts for distributed loss. The modulation frequency $\Omega$ modifies the phase mismatch, leading to a frequency-dependent transfer of energy between the FH and SH fields.
The RF-dependent phase mismatch was modeled as
\begin{equation}
\Delta k(\Omega) \;=\; \Delta k_{0} \;+\; \left(\tfrac{1}{v_{g,2\omega}} - \tfrac{1}{v_{g,\omega}}\right)\Omega,
\end{equation}
where $\Delta k_{0}$ is the residual quasi-phase mismatch at zero modulation frequency, and $v_{g,\omega}$ and $v_{g,2\omega}$ are the group velocities of the FH and SH modes, respectively. The linear dependence on $\Omega$ arises from walk-off between the FH and SH envelopes. Group velocities were obtained from Lumerical mode analysis of the waveguide cross-section, yielding $n_{g,\omega}=2.26$ and $n_{g,2\omega}=2.33$ at the FH and SH wavelengths. Propagation losses were set as $\alpha_{1}=0.2$~dB/cm for the FH and $\alpha_{2}=0.3$~dB/cm for the SH. Since the nonlinear polarization involves contributions from both waves, the effective amplitude decay was approximated as $\alpha \approx 0.25(\alpha_{1}+\alpha_{2})$.

The transfer function was normalized to $H(0)$ so that $|H(\Omega)|^2$ directly represents the modulation response relative to the static conversion efficiency. The 3-dB cutoff frequency was extracted from the normalized response $10\log_{10}|H(\Omega)|^2$. For the above parameters, the model predicts a 3-dB modulation bandwidth of approximately 147~GHz, consistent with analytical estimates based on the accumulated walk-off time~\cite{fejer2021SHGband}, $\tau = L \left| 1/v_{g,2\omega} - 1/v_{g,\omega} \right|$, which yields $f_{3\text{dB}} \approx 0.44/\tau \approx 150$~GHz.

\subsection*{Optical neuron implementation}
In the current setup shown in Fig.~\ref{fig_4}\textbf{a}, an optical neuron analogue featuring an activation-first and weighted-sum-later computational ordering is constructed to enable neuron-level all-optical operation. The fixed nonlinear activation is provided by the intensity-dependent SHG response in the PPLN waveguide, while the subsequent weighted summation for generating the output prediction signals is realized by the silicon MZI chip. This all-optical cascading experiment is designed to demonstrate single-hidden-layer, neuron-level inference (Supplementary Fig. S8) in the optical domain. With a hidden layer consisting of \(N=32\) neurons, the input vector \(\mathbf{x}\) to the optical hidden layer is first prepared through a dimensionality-mapping operation using an encoding matrix \(\mathbf{A}\) applied to the dataset feature vector \(\mathbf{x}_{\mathrm{feature}}\),

\begin{equation}
\mathbf{x} = \mathbf{A}\,\mathbf{x}_{\mathrm{feature}},
\label{eq:dim_mapping}
\end{equation}

\noindent
This dimensional alignment is performed in the electronic domain as a data pre-processing step, which is commonly adopted in integrated optical neural networks when the chip input dimension is smaller than the network dimension and when time-multiplexing schemes are employed. The elements of \(\mathbf{x}\) are subsequently normalized and mapped to the PPLN operating power range of \([0,\,0.25~\mathrm{nJ}]\) for optical signal encoding and subsequent all-optical neuron processing.
As a result, a single-hidden-layer optical network producing \(M\) scalar outputs in the cascaded PPLN--MZI system can be expressed as

\begin{equation}
y_i = \sum_{j=1}^{N} w_{ij}\, \phi_j\!\left(k_{j}\, x_j \right), 
\quad i = 1,\dots,M
\label{eq:KAN}
\end{equation}

\noindent
where \(x_j\) denotes the \(j\)-th element of the input vector \(\mathbf{x}\), \(k_{j}\) and \(w_{ij}\) are trainable coefficients, and \(\phi_j(\cdot)\) denotes the fixed univariate nonlinear activation applied to a one-dimensional projected input, which directly corresponds to the SHG-induced nonlinear mapping of optical intensity in the PPLN waveguide. This formulation of single-layer neuron operation is naturally consistent with the physical structure of our system shown in Fig.~\ref{fig_4}\textbf{a}. Specifically, the term \(k_{j}x_j\) corresponds to amplitude weighting of the input, which is encoded onto an optical carrier via a fiber-based intensity modulator. After electro–optical data carrying, the resulting signal is then injected into the PPLN chip, where the SHG process provides an element-wise nonlinear activation function \(\phi_j(\cdot)\) within the optical neuron. Finally, the outer linear combination across different hidden neurons through the coefficients \(w_{ij}\) is implemented optically via interference in the silicon MZI network , followed by temporal integration at the detector due to the limited scale of the available MZI circuitry. The resulting output $y_i$ corresponds to the prediction score used for neural inference.

Together, these steps realize an optical neuron architecture with an activation-first, weighted-summation-later ordering, which is structurally analogous to KAN-style neurons in terms of computational sequence. It should be emphasized that the nonlinear response itself is not trainable in shape; rather, the nonlinearity is provided by the fixed SHG transfer function of the PPLN waveguide. However, by introducing trainable input scaling coefficients \(k_{j}\), the operating region of this fixed nonlinear response can be effectively adjusted. This input scaling enables controlled horizontal stretching of the activation function within the accessible optical power range, thereby allowing the neuron to operate at different nonlinear regimes without altering the underlying physical nonlinearity. By jointly learning the input scaling coefficients \(k_{j}\) and the outer weights \(w_{ij}\), the network can flexibly regulate both the operating region and the strength of its nonlinear responses, which can enhance expressive efficiency when approximating structured nonlinear relationships.

\subsection*{Digital training procedure}
All models are trained using Python 3.9 with the PyTorch open-sourced framework, and training is accelerated using a CUDA-enabled graphics processing unit (NVIDIA RTX 4090, 24GB RAM). 
Each model is trained with task-specified epochs using the Adam optimizer with an initial learning rate of $10^{-3}$ and $L_2$ regularization to mitigate overfitting. Cross-entropy loss is employed as the objective function, with a batch size of 128. A learning rate scheduler reduces the learning rate by a factor of 10 after 40\% and 80\% of the total training epochs to facilitate convergence.
To enhance generalization in the medical image classification task, dropout is applied to the hidden layers with a rate of 0.2. Additionally, batch normalisation is employed after each linear transformation to stabilise training and improve gradient flow.

\subsection*{Linear weight mapping}
The obtained weight matrix after training for each optical neuron is partitioned into 2×2 submatrices and mapped onto the fabricated silicon photonics-based universal linear processor. The \textit{i}-th submatrix $A_i$ is decomposed via singular value decomposition as $A_i=U_i\Sigma_iV_i$, where $U_i$ and $V_i$ are unitary matrices and $\Sigma_i$ is a diagonal matrix. Following the Clements design, diagonal matrices are implemented using arrays of amplitude phase shifters with transfer matrices of the form $D = \mathrm{diag}(e^{i\phi_1}, e^{i\phi_2})$, while unitary matrices are realized using MZIs, whose transfer matrix is given by the following:

\begin{equation}
\label{eq2}
T = i e^{\frac{i\theta}{2}} 
\begin{bmatrix}
e^{i\phi_3} \sin\left(\frac{\theta}{2}\right) & e^{i\phi_3} \cos\left(\frac{\theta}{2}\right) \\
\cos\left(\frac{\theta}{2}\right) & -\sin\left(\frac{\theta}{2}\right)
\end{bmatrix}
\end{equation}

\noindent
The reprogrammability of the processor is facilitated by precise tuning of both internal and external MZI phase shifts, denoted by $\theta$ and $\phi_i$. These phase shifts are adjusted using resistive heaters based on strong thermo-optic effect in silicon. 

\begin{equation}
\label{eq2.4}
\Delta \Phi = \left( \frac{2\pi L}{\lambda_0} \right) \left( \frac{dn}{dT} \right) \varepsilon V^2
\end{equation}

\noindent
Here, $L = 100$ \textmu m denotes the length of the heater, $\lambda_0=1550$ nm is the vacuum wavelength of the on-chip coupled light, and the thermo-optic coefficient of silicon at 300 K is $\frac{dn}{dT} = 1.86 \times 10^{-4}~\text{K}^{-1}$. The constant $\varepsilon$ accounts for thermal properties including heat capacity and resistance. The phase shift induced by a single heater exhibits a square-law dependence on the applied voltage. In the proposed optical neuron, both inputs and weight matrices are encoded through voltage-controlled phase shifts. When optical signals are injected, the silicon photonic MZI system performs the corresponding linear operations in the optical domain.

\section*{Acknowledgements}
This research was funded by the NUS Artificial Intelligence Institute (NAII) seed grant number NAII-SF-2024-001.

\section*{Competing interests}
The authors declare no competing interests.

\section*{Availability of data and materials}
The data that support the findings of this study are available from the corresponding author upon reasonable request.

\section*{Author contributions}
W.F. and X.S. contributed equally to this work. W.F., X.S., D.Z., and A.D. conceived the idea and co-wrote the manuscript. W.F., X.S., S.S.M., Z.W., Y.G., J.W. and X.C. designed the experiments and carried out device fabrication. W.F., X.S. and L.S. developed the simulations. W.F. and Z.W. conducted the literature review. A.D. and D.Z. supervised the project. All authors contributed to the discussion of results and completion of the manuscript.

\bibliography{refs}

\end{document}